\documentclass[aps,prb,showpacs,preprint]{revtex4}
\usepackage{graphicx}
\input{epsf}

\begin{document}
\title {Resonant enhancement of MQT in Josephson junctions: the influence of coherent two-level systems}
\author{M. V. Fistul$^{1,2}$}

\affiliation {$^{1}$ Theoretische Physik III, Ruhr-Universit\"at
Bochum, D-44801 Bochum, Germany \\
$^{2}$ National University of Science and Technology MISIS, Moscow 119049, Russia }

\date{\today}
\begin{abstract}
We report a theoretical study of the macroscopic quantum tunneling (MQT) in small Josephson junctions containing randomly distributed two-level systems. We focus on  the magnetic field dependent crossover temperature $T_{cr}$ between the thermal fluctuation and quantum regimes of switching from the superconducting (the zero-voltage) state to a resistive one. In the absence of two-levels systems the crossover temperature shows a smooth decrease with an applied magnetic field characterized by an external flux $\Phi$. Beyond that  we predict a narrow peak  in the dependence of $T_{cr}(\Phi)$ occurring in the intermediate range of $\Phi$.
The effect becomes more pronounced as the junction size increases.
We explain this effect quantitatively by a strong resonant suppression of a potential barrier for the Josephson phase escape that is due to  the \emph{coherent quantum Rabi oscillations} in two-level systems present in the junction.
\end{abstract}


\maketitle
\section{Introduction}

Great interest is currently attracted to experimental and theoretical studies of macroscopic quantum phenomena in diverse
Josephson systems \cite{Tinkham,qubits,Legg,Clarke1,Martinis-Lukens-et-al,MQPh1,MQPh2,MQTHTSC1,Fglstate}. It is a well known that at low temperatures
the switching from the superconducting (the zero-voltage) state to a resistive one occurs in the form
of macroscopic quantum tunneling (MQT) of a Josephson phase \cite{Tinkham,Legg,Clarke1,MQPh1,MQPh2,MQTHTSC1,Fglstate}. At high temperatures this so-called Josephson phase
escape phenomenon is determined by thermal fluctuations. The crossover temperature $T_{cr}$ between these two regimes in a simplest case of a single
degree of freedom, i.e. the Josephson phase $\varphi$, is determined by the frequency of small oscillations $\omega_0$ of the Josephson phase
on the bottom of potential well. Since the Josephson phase plasma frequency $\omega_p$ is determined by the critical current $I_c$  as $\omega_p~\propto~I_c^{1/2}$, one can expect
that $\omega_0$ and the crossover temperature $T_{cr}$ vary with magnetic field. Indeed, the crossover temperature is written as \cite{Tinkham}
$
k_BT_{cr}=\frac{\hbar \omega_0}{2\pi}=\frac{\hbar \omega_p}{2\pi}(1-j)^{1/4},
$
where $j=I/I_c$ is the normalized external current $I$. The typical values of the external current $I$ can be evaluated from the condition that the Josephson phase escape in the MQT regime occurs as
the potential barrier $U_0~\simeq~\frac{\hbar I_c}{2e}(1-j)^{3/2}$ becomes comparable with the energy of small oscillations $\hbar \omega_0$, i.e.  $(1-j)~\propto~I_c^{-2/5}$, and we obtain
\begin{equation} \label{Cross-Temp-SA-main}
k_BT_{cr}~\propto~I_c^{2/5}~.
\end{equation}

Since the applied magnetic field characterized by an external magnetic flux $\Phi$, results in the reduction of the critical current $I_c$ one can expect a smooth decrease of the crossover temperature $T_{cr}$ with $\Phi$.


A crucial condition allowing one to obtain the dependence (\ref{Cross-Temp-SA-main}) is the absence of interactions of the Josephson phase
with other degrees of freedom. E.g. one can expect in the correspondence with a generic analysis \cite{Legg} that the Josephson phase interaction with a large amount of linear oscillators (such an interaction has been used as a model of dissipative  environment) results in a suppression of both the MQT and the crossover temperature. However, a careful preparation of experimental setup has allowed one to reduce these undesirable effects.

The interaction of a Josephson phase with other degrees of freedom can result in the \emph{reduction} of the potential barrier, and therefore, lead to an enhancement of the MQT \cite{MQTHTSC1,Fglstate,FistWalUst,VarOvch,FistUst-CM}. E.g. in the presence of magnetic field the intrinsic cavity modes  are excited, and an enhancement of the MQT has been obtained \cite{VarOvch,FistUst-CM}. However, for small junctions the probability of the cavity modes excitation becomes rather small \cite{VarOvch}, and an enhancement of MQT is also small. Moreover, such an enhancement of MQT has to lead to  a smooth dependence of the crossover temperature $T_{cr}$ on an applied magnetic field. In all these cases the equilibrium state of a Josephson junction in which  the Josephson phase interacts with a set of \emph{linear oscillators}, has been considered.

Notice here that a strong back influence of the Josephson phase on the oscillators dynamics suppresses the resonant effects, and therefore, the interaction of the Josephson phase with a set of linear oscillators can not lead to the resonant enhancement of the MQT.

On other hand, we also recall that in the \emph{non-equilibrium case} as the Josephson junction is subject to an externally applied microwave radiation of the frequency $\omega$, the extremely pronounced resonant effects have been observed in the MQT phenomena \cite{FistWalUst,ACdrivenJJ}. Such a resonant interaction of the Josephson phase with an applied microwave radiation results in a strong suppression of the potential barrier, and therefore, to the resonant enhancement of the MQT as $\omega_0~\simeq~\omega$.

Therefore, an interesting question  arises in this field: is it possible to observe the resonant enhancement of the MQT in the equilibrium state of the Josephson junction?

In this paper we show that such a resonant enhancement of the MQT naturally occurs in the Josephson junctions containing a large amount of microscopic coherent two-levels systems (TLSs). Indeed,
it is well known that the TLSs so-called "fluctuators" are intrinsically present in a Josephson junction with amorphous interlayer \cite{Martinis,Ustinov-TLS-1,Ustinov-TLS-2}. These defects can exist in two quantum states with an energy separation between them  $\Delta_0$ and the tunneling splitting $\Delta$. At high temperatures $k_BT \geq \Delta$ or as $\Delta \leq \Delta_0$  these centers display a random (the Poisson noise) but in the opposite regime ($k_BT<\Delta$ and $\Delta>\Delta_0$ ) TLSs establish the coherent quantum oscillations (Rabi oscillations) of the frequency $\Omega=\Delta/\hbar$. Thus, one can expect that these TLSs resonantly excite the oscillations of the Josephson phase as the magnetic field and bias current dependent frequency of small oscillations of the Josephson phase $\omega_0$ matches the frequency of intrinsic quantum oscillations $\Omega$. Moreover, the back influence of the Josephson phase dynamics on the dynamics of coherent two-levels systems is rather small in the usual regime as the tunneling splitting between the two low-lying states, i.e. $\Delta$ , is much smaller than other energy differences.  These oscillations, similarly to a non-equilibrium case \cite{FistWalUst}, can result in a strong reduction of the potential barrier and the enhancement of the MQT.

Below we present a model and the Josephson phase dynamics in the presence of coherent two-levels defects (Section II),  and the quantitative analysis of the resonant enhancement of the MQT (the Section III). The Section IV provides conclusions.


\section{The Josephson junction containing TLS in the presence of magnetic field: the Josephson phase dynamics.}

Next, we  quantitatively analyze the MQT regime of the Josephson phase escape in small  (the size $L<\lambda_J$, where $\lambda_J$  is the Josephson penetration length) Josephson junctions in the presence of magnetic field. We also take  into account a weak dipole-dipole  interaction of the Josephson phase with the coherent TLSs present in the junction.
A small Josephson junction subject to an externally applied magnetic field is characterized by time $t$ and coordinate $x$ dependent Josephson phase $\varphi(t,x)$.
In the presence of magnetic field applied in the junction plane and directed perpendicular to the $x$ axis, the Josephson phase is written as
\begin{equation} \label{JP-MagField}
\varphi(t,x)~=~\frac{2\pi \Phi x}{\phi_0 L}
+\chi(t,x)~~,
\end{equation}
where $\Phi$ is the external magnetic flux, and $\phi_0=hc/2e$ is
the flux quantum.

For small Josephson junction we can neglect the excitation of cavity modes, and the Hamiltonian $H_J$ depends on  spatially averaged
time-dependent Josephson phase $\chi(t)$ as
\begin{equation} \label{Hamilt:total}
H_J ~=E_{J0} \Biggl [
\frac{1}{2\omega_{p0}^2}\dot{\chi}^2+\gamma(\Phi)[1-cos(\chi)]-j\chi \Biggr ]~,
\end{equation}
where the magnetic field dependent parameter $\gamma(\Phi)$ determines the reduction of the junction critical current with the magnetic field as
\begin{equation} \label{CriticalCurr-NoMagnField}
I_c(\Phi)=I_{c0}\gamma(\Phi)=I_{c0}|\frac{\phi_0}{\pi \Phi}\sin(\frac{\pi \Phi}{\phi_0})|.
\end{equation}
The typical dependence of the Josephson current on the magnetic field is shown in Fig. 1.
\begin{figure}[tbp]
\includegraphics[width=\columnwidth,angle=0]{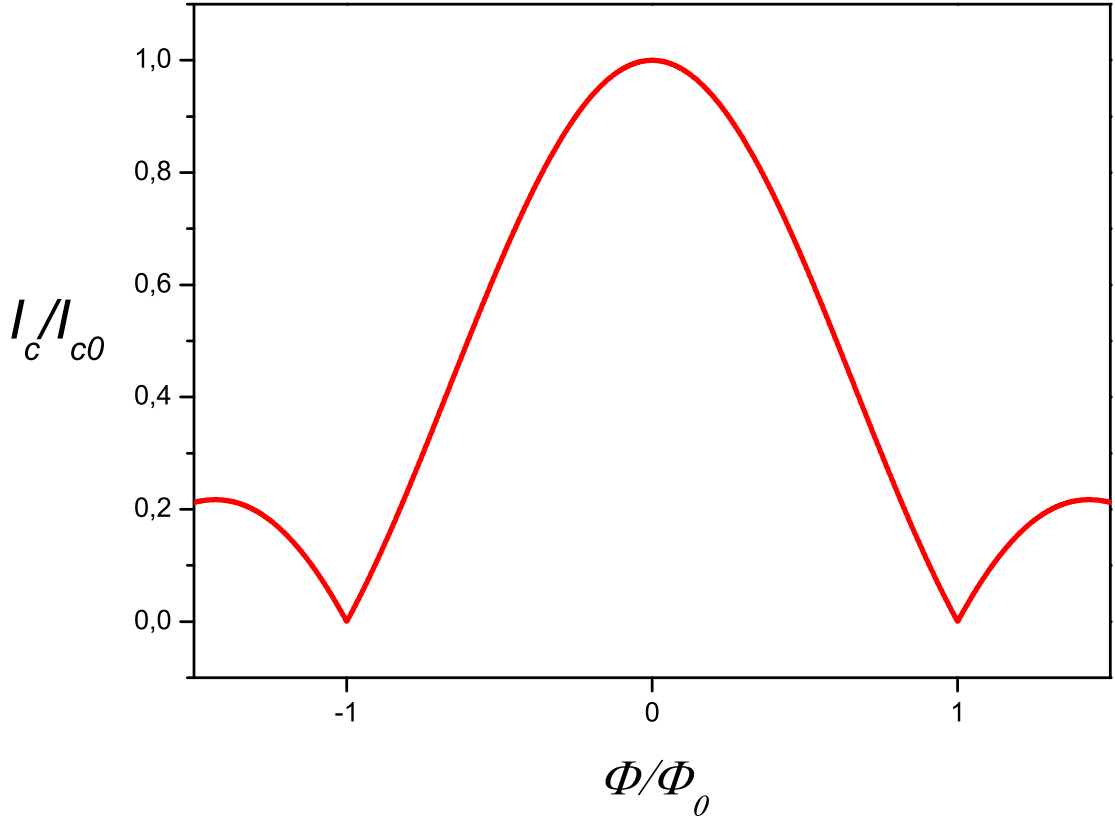}
\caption{\label{criticalcurrent} (Color online) The typical dependence of the Josephson critical current on the magnetic field for small Josephson junctions [Eq. (\ref{CriticalCurr-NoMagnField})].}
\end{figure}
Here,  $E_{J0}$ and $\omega_{p0}$ are the Josephson coupling energy and the Josephson plasma frequency in the absence of magnetic field, accordingly. The normalized dc
bias $j=I/I_{c0}$ allows one to effectively tune (decrease) the potential relief [see, the Eq. (\ref{Hamilt:total})] for the Josephson phase.

We also take into account the quantum two-levels defects distributed in the insulator layer of the Josephson junction. The Hamiltonian of such TLSs reads as
\begin{equation} \label{Hamilt-TLS}
H_{TLS}=\sum_i \frac{m}{2}[\dot{\Psi}_i]^2+U(\Psi_i)~,
\end{equation}
where $\Psi_i$ and $U(\Psi_i)$ are correspondingly the degree of freedom and a double-well potential characterizing a single TLS; $m$ is the effective mass of the TLSs. The Hamiltonian of
a weak dipole-dipole interaction between TLSs and the Josephson phase is written in the following form:
\begin{equation} \label{Hamilt-TLS-int}
H_{int}=\frac{E_{J0}\eta}{\omega_{p0}^2}\sum_i \ddot{\Psi}_i(t) \chi(t)~,
\end{equation}
where the dimensionless parameter $\eta$ determines the interaction strength of the Josephson phase with a single TLS.
So the total Hamiltonian of the Josephson junction containing the TLSs in the presence of magnetic field is written as
\begin{equation} \label{total-Hamiltonian}
H=H_J\{\chi\}+H_{TLS}+H_{int}~.
\end{equation}

The dynamics of the Josephson phase $\chi(t)$ interacting with  the TLSs is described by the following (nonlinear) differential equation
\begin{equation} \label{Equation-saddle}
-\frac{1}{{\omega_{p0}}^2}\ddot{\chi}(t)+V^\prime (\chi)=\frac{\eta}{\omega_{p0}^2} \sum_i \ddot{\Psi}_i (t)~,
\end{equation}
where the potential $V(\chi)=\gamma(\Phi)[1-cos(\chi)]-j\chi$.

To solve this equation we represent the Josephson phase $\chi(t)$ as a sum of low- and high frequency terms, i.e.
$\chi(t)=\frac{\pi}{2}+\chi_0(t)+\xi(t)$. The high-frequency term is found as
\begin{equation} \label{Highfr}
\xi(t)=\frac{\eta}{\omega_{p0}^2}\sum_i\int_0^{\infty} d\tau_1 G(t-t_1)\ddot{\Psi}_i(t_1),
\end{equation}
where $G(t)$ is the Green function of the linearized equation (\ref{Equation-saddle}) that reads as:
\begin{equation} \label{Linearequation}
-\frac{1}{\omega_{p0}^2}\ddot{Y}(t)+\gamma(\Phi)\chi_0Y(t)=0~.
\end{equation}
Explicitly the Green function $G(t)$ is written as
\begin{equation} \label{Linearequation-2}
G(t)=\int \frac{d\omega}{2\pi} G(\omega) e^{i\omega t}=\int \frac{d\omega}{2\pi} e^{i\omega t}\frac{\omega_{p0}^2}{\omega^2-\omega_{p0}^2 \gamma(\Phi)\chi_0+i\alpha \omega_{p0}} ~.
\end{equation}
Here, we introduce the parameter $\alpha$ in order to describe the dissipative effects in the Josephson phase dynamics.

Notice here, that in Eqs. (\ref{Highfr}) and (\ref{Linearequation-2}) we neglected the term that is due to the back influence of the Josephson dynamics on the coherent quantum oscillations of TLSs. That is valid as two low-lying states are well separated from other energy levels. In this case, the all quantum degrees of freedom, $\Psi_i$,  do not depend on $\chi(t)$, and TLSs just show the quantum Rabi oscillations with the frequencies $\Delta_i$ in the ground state.

Due to the nonlinearity of the potential $V(\chi)$ the high frequency term $\xi(t)$ leads to the effective reduction of this potential as
\begin{equation} \label{Effpotential}
V_{eff}=V(\chi_0)-\frac{\xi^2(\chi_0)}{2} \chi_0~.
\end{equation}
The equilibrium value of $\chi_0$ is determined by an external bias current as
\begin{equation} \label{Chi-0}
\chi_0=\sqrt{2\delta}=\sqrt{\frac{2[j-\gamma(\Phi))]}{\gamma(\Phi)}}.
\end{equation}

\section{The MQT and crossover temperature $T_{cr}$: influence of Rabi oscillations in the ground state of TLS}

By making use of a standard analysis of the MQT \cite{Legg,Col} we obtain the escape rate of Josephson phase in the MQT regime as:
\begin{equation} \label{EscapeRate}
\Gamma(\delta)~\simeq~\exp\{-\frac{36E_{J0}\gamma^{1/2}(\Phi)}{5\hbar \omega_{p0}}\{ 2[\delta-\xi^2(\chi_0)/2)]\}^{5/4}\}~~.
\end{equation}
Taking into account that the typical values of $\delta=\delta_0$ allowing non-negligible escape of the Josephson phase, are determined by the condition $\Gamma (\delta)~\simeq~1$ we obtain the transcendent equation for $\delta_0$
\begin{equation} \label{Delta0}
\delta_0-\xi^2(\delta_0)/2 \simeq~(\frac{\hbar \omega_{p0}}{E_{J0}})^{4/5}\gamma^{-2/5}(\Phi)
\end{equation}
and the crossover temperature $T_{cr}$ is determined by equation
\begin{equation} \label{CrossTemp}
T_{cr}=\frac{\hbar \omega_{p0} \gamma^{1/2}(\Phi)}{2\pi k_B                                                                      }(2\delta_0)^{1/4}~~.
\end{equation}
 Since in the absence of TLSs  the parameter $\xi=0$  we arrive on the Eq. (\ref{Cross-Temp-SA-main}) for the magnetic field dependence of $T_{cr}$.
 It reads explicitly as
\begin{equation} \label{CrossTemp-Expl}
T_{cr}=\frac{\hbar \omega_{p0}}{2 \pi k_B}(\frac{\hbar \omega_{p0}}{E_{J0}})^{1/5}\gamma^{2/5}(\Phi).
\end{equation}
Such a smooth dependence is shown in Fig. 2 by  blue (black) line. Next, we notice that as the Josephson phase weakly interacts with a set of linear oscillators the deviation from the Eq. (\ref{Cross-Temp-SA-main}) is also small because in this case the function $G(t)$ does not display the resonant behavior.

The situation changes drastically as we turn to the interaction of the Josephson phase with the coherent TLSs. It is well known that in the ground state the TLSs display coherent quantum Rabi oscillations, and quantitatively the coherent dynamics of TLSs is described by Bloch equations \cite{LeggettRMP}.
In this case we obtain
\begin{equation} \label{Ksi}
\xi^2=\frac{\eta^2}{\omega_{p0}^4}\sum_i\int dt_1 \int dt_2 \ddot{G}(t-t_1)\ddot{G}(t-t_2)K_i(t_1-t_2)~~,
\end{equation}
where the time-dependent correlation function $K_i(t_1-t_2)=<\Psi_i(t_1)\Psi_i(t_2)>$ of coherent TLSs has been obtained in Refs. \cite{ChackKivelson,LeggettRMP,FistVolk} as
\begin{equation} \label{CorrFunction}
K_i^0(t)= \Psi_0^2 e^{-2i\Delta t/\hbar-\tilde{\Gamma}|t|}~,
\end{equation}
where  $\pm \Psi_0$ are the values of $\Psi$ on the bottom of potential wells, the $\Delta_i$ is the energy splitting of $i$-th TLS, and the parameter $\tilde{\Gamma}$ describes the decay rate of quantum Rabi oscillations.
Substituting (\ref{CorrFunction}) in  (\ref{Ksi}) we obtain that the parameter $\xi$ has a resonant form as
\begin{equation} \label{xi-Fin}
\xi=\eta^2 \Psi_0^2 \sum_i\Delta_i^2[(\Delta_{i}-\hbar \omega_{p0} \gamma^{1/2}(\Phi)(2\delta_0)^{1/4})^2+\tilde{\alpha}^2 (\hbar \omega_{p0})^2]^{-1}.
\end{equation}
Here, the parameter $\tilde{\alpha}$ describes all dissipative effects in both the TLSs and Josephson phase dynamics.

Next analysis depends on the distribution of tunneling splitting energies $\Delta_i$ of TLSs in Josephson junctions. First, for simplicity we consider a case where all TLSs have the same splitting energies $\Delta$. In this case all TLSs give the same contribution to $\xi^2$, and substituting (\ref{xi-Fin}) in (\ref{Delta0}) and (\ref{CrossTemp}) we obtain the resonant type of enhancement of MQT that is due to the interaction of the Josephson phase with TLSs. The typical dependency  of $T_{cr}$ on the magnetic field is shown in Fig. 2 by red (gray) line.
\begin{figure}[tbp]
\includegraphics[width=\columnwidth,angle=0]{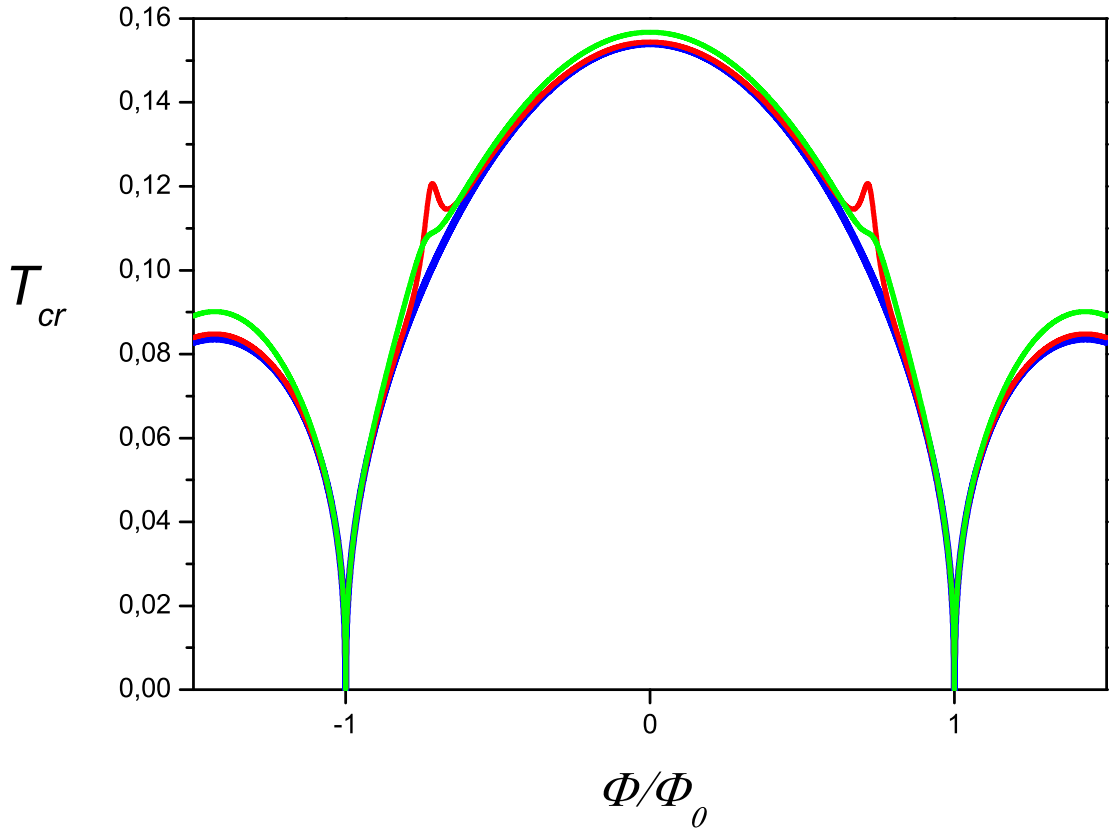}
\caption{\label{crossovertemp} (Color online) The typical dependencies of the crossover temperature $T_{cr}$ on the magnetic field: the interaction with TLSs is absent (the blue (black) line); a weak interaction with the TLSs having the same splitting energies $\Delta$ (the red (gray) line); a weak interaction with the TLSs having randomly distributed splitting energies in the range between $\Delta_{max}$ and $\Delta_{min}$ (the green (light gray) line). The parameters
$\Delta/(2\pi k_B)=\Delta_{max}/(2\pi k_B)=0.1 K$ and $\Delta_{min}/(2\pi k_B)=0.02 K $ has been chosen. The quality factor of the resonance is equal $20$. }
\end{figure}
In a more realistic case as  a tunneling splitting $\Delta$ is randomly distributed in a wide range, i.e. $\Delta_{min}<\Delta<\Delta_{max}$, the enhancement of MQT occurs also in a wide range of  magnetic field $\Phi$. The typical dependency of $T_{cr}$ on the magnetic field for this case is shown in Fig. 2 by  green (light gray) line.

\section{Conclusion}
We theoretically studied the macroscopic quantum tunneling (MQT) phenomenon in the Josephson junction containing a large amount of quantum two-levels systems (TLSs). The MQT is characterized by the crossover temperature $T_{cr}$ between the thermal fluctuations and quantum tunneling regimes of the Josephson phase escape. We focused on the dependence of $T_{cr}$ on an externally applied magnetic field characterizing by magnetic flux $\Phi$.
In the absence of TLSs the crossover temperature $T_{cr}$ displays a smooth decrease with $\Phi$  in the range $-\phi_0< \Phi <\phi_0$ [see, Eqs. (\ref{Cross-Temp-SA-main}) and  (\ref{CrossTemp-Expl}), the typical dependence is shown in Fig. 2 by blue (black) line].

In the presence of even a weak interaction of the Josephson phase with the TLSs we predicted a \emph{resonant enhancement } of the MQT. Such an enhancement is explained by a resonant suppression of the potential barrier for the Josephson phase escape which, in turn, due to the presence of the coherent Rabi oscillations in the ground state of coherent quantum TLSs. Such an effect resembles a resonant suppression of the potential barrier in the Josephson junctions subject to an externally applied microwave radiation \cite{FistWalUst,ACdrivenJJ}. This enhancement is especially strong if the TLSs have the same splitting energies $\Delta$. The typical dependence of $T_{cr}(\Phi)$ for this case is shown in Fig. 2 by red (gray) line.
In an opposite case as the tunneling splitting displays a wide distribution we obtain that the enhancement of the MQT occurs in a wide range of magnetic fields [see, the Fig. 2 green (light gray) line].

Finally, we notice that a crucial condition allowing one to observe such a resonant enhancement of the MQT is the presence in the Josephson junctions a large amount of coherent TLSs, i.e. the TLSs with the tunneling splitting $\Delta ~\simeq~k_B T$, and $\Delta \gg \Delta_0$, where $\Delta_0$ is the energy difference between the minimums of a double-well potential characterizing the TLSs. The quality factors of both the Josephson junction and the TLSs have to be large.

\textbf{Acknowledgement}
We would like to acknowledge Alexey V. Ustinov for fruitful discussions.
We acknowledge the financial support of the Ministry of Education and Science of the Russian Federation  in the framework of Increase Competitiveness Program of NUST "MISiS"($K2-2014-015$).

\end{document}